\documentclass[sigconf, screen]{acmart}

\usepackage{multirow} 
\usepackage{enumitem}
\usepackage[table]{xcolor} 

\definecolor{green}{RGB}{200, 230, 200} 
\definecolor{red}{RGB}{255, 220, 220} 
\definecolor{blue}{RGB}{200,220,255} 
\AtBeginDocument{%
  }

\setcopyright{acmlicensed}
\copyrightyear{2025}
\acmYear{2025}
\acmDOI{XXXXXXX.XXXXXXX}
\acmConference[KDD '25]{31st ACM SIGKDD Conference on Knowledge Discovery and Data Mining (KDD 2025) Workshop on End-to-End Customer Journey Optimization}{August 03--07, 2025}{Toronto, ON, Canada}
\acmISBN{978-1-4503-XXXX-X/18/06}

\begin{document}

\title{PCGNet: Unifying Shared and Specific Information for Fashion Matching Recommendations}

\author{Shuiying Liao}
\affiliation{
  \institution{The Hong Kong University of Science and Technology}
  \country{Hong Kong}}
\email{shuiyingl@ust.hk}

\author{P.~Y.~Mok*}
\affiliation{
  \institution{The Hong Kong University of Science and Technology}
  \country{Hong Kong}}
\email{tracy.mok@ust.hk}

\renewcommand{\shortauthors}{Liao et al.}

\begin{abstract}
With the continuous evolution of e-commerce platforms, enhancing user experience and customer satisfaction are of paramount importance. In the fashion domain, recommending complementary clothing items that match selected pieces is a crucial cross-selling technique that improves customer satisfaction. Nevertheless, fashion matching presents significant challenges, as recommendations must not only align with individual users’ fashion preferences but also ensure compatibility between garments. These challenges are twofold. First, existing models often assume an overly simplified decoupled relationship between product compatibility and personalized user preferences, overlooking the natural complexity between the two. Second, existing data-driven approaches are not optimized for real-world fashion data, which is typically sparse and characterized by noisy interactions. To address these challenges, we propose \textit{Personalized Compatibility Graph Network} (\textbf{PCGNet}), a multi-objective graph learning framework that organically unifies the modeling of product compatibility and personal preferences. PCGNet empolys contrastive mutual information maximization to extract and align shared and view-specific patterns, thereby capturing the complex interplay between compatibility and personal preferences. Moreover, we introduce a correlation-aware neighbor sampling and a learnable global graph augmentation, which enhance the model by incorporating self-supervised signals mined directly from the graph, ensuring more stable and informative representations. Finally, PCGNet generates recommendation scores through the joint optimization of BPR ranking loss and multi-view mutual information losses. Experimental validation on two benchmark datasets demonstrates that PCGNet significantly outperforming state-of-the-art methods across all four evaluation metrics. By enabling more effective representation learning, PCGNet offers novel insights into future customer behavior prediction research. 
\end{abstract}




% \received{20 February 2007}
% \received[revised]{12 March 2009}
% \received[accepted]{5 June 2009}

\maketitle

\section{Introduction}
With the rapid growth of online fashion platforms, improving user experience and satisfaction have become essential for the survival of these shopping platforms. By leveraging techniques such as self-supervised learning and multi-objective learning, it is now possible to gain deeper insight into customer behavior and improve personalized recommendation strategies. As a result, online fashion platforms have increasingly demanded intelligent clothing recommendation systems. Unlike traditional recommendation tasks, fashion matching recommendation requires a dual focus on (1) users' personalized preferences, referred to as \textbf{personalization} \cite{GPBPR}, and (2) the visual and descriptive attributes of clothing products to ensure \textbf{compatibility} between matching pairs (e.g., top and bottom garments) \cite{zhou2022attribute}. These dual focuses introduce unique challenges in data science and recommendation research~\cite{VBPR, GPBPR, liao2023, liao2025data, liao2026consistency}.

\begin{figure}[t]
  \centering
  \vspace{-5px}
  \setlength{\abovecaptionskip}{0.2cm}
  \includegraphics[width=0.95\linewidth]{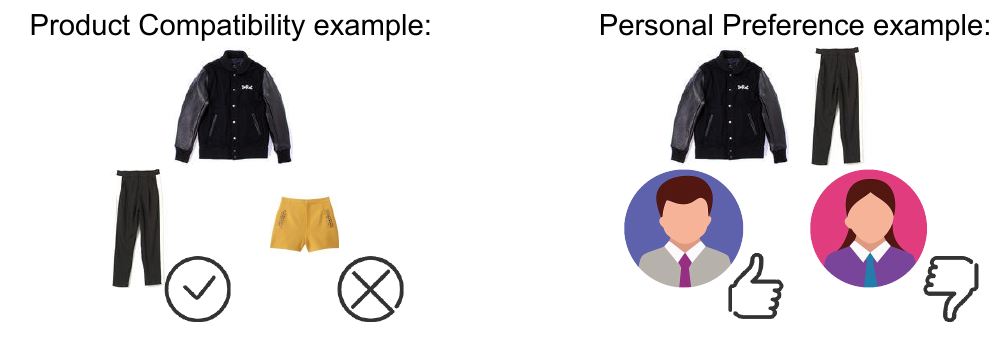}
  \caption{Personalized fashion matching examples for various users.}
  \label{fig: CP_example}
  \vspace{-10px}
\end{figure}

Traditional fashion matching recommendation methods suffer from two key limitations. 
First, previous work either treats product compatibility and personal preference as isolated objectives \cite{VBPR, lu2021outfit, li2022disentangled} or combines them in a linearly and decoupled manner~\cite{GPBPR}.
Preference-centric methods \cite{BPR, VBPR} predict user-item affinity without considering outfit compatibility, while compatibility-centric methods \cite{li2017mining, han2017learning, cui2019dressing, vivek2023personalized, jing2023category, guan2022partially, liao2025data, liao2026hamiltonian, liao2024reproducibility} model item-item relationships (e.g., color coordination) and align multi-modal features in latent spaces but fail to consider user-specific tastes for recommendable matching garments. These approaches are later combined through post-hoc fusion \cite{GPBPR, 
mo2023towards, PCE}, yet such decoupling fails to capture their complex context-dependent interplay.
In this paper, we challenge the existing approaches of modeling personalization and compatibility as mutually exclusive factors. In other words, we reject the assumption that a high level of personalization must necessarily result in a low level compatibility, or vice versa. 
For instance, as illustrated in Fig.~\ref{fig: CP_example}, aesthetically matched outfits may not be universally accepted by all users, even though a general sense of aesthetics is derived from the choices of many users.
Furthermore, existing multi-view frameworks lack mechanisms to dynamically align shared and view-specific signals, leading to suboptimal recommendations. Although attention-based fusion \cite{liu2024unifying} and hierarchical frameworks \cite{CP} improved flexibility, they still relied on manually designed interaction weights.

The second key limitation is that traditional graph-based recommendation systems often suffer from sparsity. User preferences and compatible item pairs are often inferred from implicit feedback (e.g., clicks), which is noisy and incomplete. Some studies~\cite{graphsurvey, fan2019graph, zhang2023constrained} 
directly adopt raw interaction data to construct graphs, ignoring the inherent noise and task-irrelevant connections in real-world fashion datasets. This can compromise the robustness of the resulting model and even propagate misleading information during graph convolution \cite{jin2020graph}.

\begin{figure}[t]
  \centering
  \vspace{-8px}
  \setlength{\abovecaptionskip}{0.2cm}
  \includegraphics[width=0.95\linewidth]{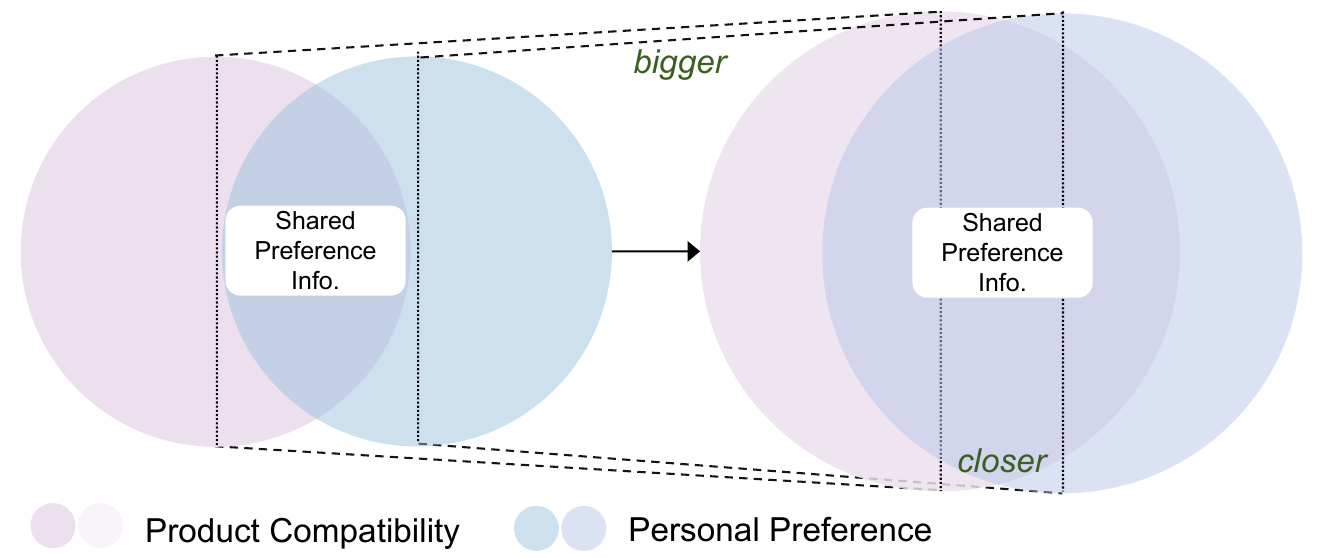}
  \caption{Conceptual illustration of our motivation: coupled overlap (shared information) and divergence (view-specific information) between compatibility and preference.}
  \label{fig: motivation}
  % \vspace{-15px}
\end{figure}

To address these challenges, we introduce the \textit{\textbf{Personalized Compatibility Graph Network}} (\textbf{PCGNet}) -- a pioneering framework that unifies the modeling of shared information alongside the product-compatibility and personal-preference specific information. This is achieved through a \textbf{mutual information-based multi-view heterogeneous graph learning framework}, which substantially improves the reliability of the graph structure. 
PCGNet is built on three core modules: 
(1) \textbf{Dual-view heterogenrous graph} -- we construct a product-compatibility graph that captures the intrinsic matching rules between fashion product pairs, and a personal-preference graph, which models individual tastes. By maximizing cross-view mutual information, we extract shared features while preserving view-specific information. (See a concept illustration in Fig.~\ref{fig: motivation}.) For instance, product-compatibility specific information may include color coordination rules, whereas personal-preference specific information could involve scene-sensitive patterns. 
(2) \textbf{Correlation-aware global neighbor sampling} -- we introduce a dynamic filtering strategy that selects high-confidence edges (interactions), ensuring the preservation of compatibility and personal-preference relationships
based on node embedding similarity~\cite{liu2022rgcf}. Moreover, we incorporate a \textbf{learnable graph augmentation generator}, utilizing the Gumbel reparameterization trick~\cite{maddison2016concrete, jang2016categorical, 2024beyond} to create augmented graphs, effectively capturing view-specific information and mitigating deficiencies caused by data sparsity.
(3) \textbf{Attentive graph fusion module and multi-objective optimization} -- We integrate information from both views through an attentive graph fusion module and a shared graph encoder. Furthermore, we design a multi-objective function, incorporating BPR loss, shared information loss, and specific information loss, to optimize both product-compatibility and personal-preference simultaneously.

% 1) Cross-view representation learning from heterogeneous graphs, 
% 2) Robust correlation-aware relevance-guided neighbor sampling strategy, to dynamically filter task-irrelevant edges while preserving compatibility and preference relationships
% %Robust graph structure learning via subgraph sampling 
% \cite{liu2022rgcf}, and 3) By maximizing shared and unique information between heterogeneous graphs from different views for multi-target alignment \cite{velivckovic2018deep, 2024beyond}, PCGNet disentangles shared and unique signals across compatibility and preference views, enabling finer-grained user-item relationship modeling.

Our contributions are summarized as follows:
\begin{itemize}

% \item We propose PCGNet, the first framework to unify compatibility modeling and personalized preference learning through dual-graph mutual information maximization, addressing their complex interactions.

\item We redefine the modeling criteria for fashion matching recommendations and design PCGNet -- a framework that dynamically aligns product compatibility and personal preference through mutual information maximization. This approach captures both their non-linear overlap (shared information) and divergence (view-specific patterns). PCGNet jointly optimizes BPR loss (supervised ranking) and cross-view mutual information (self-supervised alignment), ensuring multi-objective synergy.

% Our proposed multi-task objective not only maximizes the overlap between product-compatibility and personal preferences but also ensures that each aspect retains a sufficient amount of view-specific information.
% \item We design a relevance-guided sampling mechanism to construct noise-resistant subgraphs, significantly improving the robustness of graph-based fashion recommendations.
\item We refines raw noisy graph using correlation-aware neighbor sampling and learnable augmentation, leveraging both labeled and unlabeled data for robust representation learning. 
% \item A correlation-aware sampling strategy with a learnable graph augmentation generator are utilized to optimize the graph structure, ensuring robustness and improving recommendation accuracy through joint optimization.

\item Extensive experiments on tow benchmark public datasets demonstrate PCGNet’s superiority over state-of-the-art baselines. Ablation studies further validate the necessity and effectiveness of each component of the PCGNet.
\end{itemize}

\section{Related Work}
\subsection{Personalized fashion compatibility modeling}
Personalized fashion compatibility modeling aims to simultaneously maximize \textit{visual/textual outfit compatibility} while satisfying \textit{user-specific preferences}, attracting growing attention in both academia and the fashion industry. Early approaches primarily focused on visual/textual compatibility through metric learning. For example, VBPR \cite{VBPR} integrated visual features from CNNs with Bayesian personalized ranking (BPR) to model personal preferences, while \citet{jing2023multimodal} proposed type-aware embeddings to capture category-level compatibility rules (e.g., `skirts pair well with blouses'). Subsequent work introduced hierarchical frameworks, learning compatibility at both item and outfit levels using graph convolutional networks (GCNs) \cite{ECCV2018}.  
Recent endeavors emphasized \textit{joint modeling of compatibility and preferences}. \citet{lu2021outfit} designed a multi-layered matching network that combines compatibility scores with user historical interactions, but its linear and decoupled fusion mechanism fails to capture contextual dependencies. \citet{li2022disentangled} employed variational autoencoders to disentangle user intent into compatibility and preference factors, yet their static disentanglement overlooks dynamic scenario-based interactions, for example, a user might prioritize compatibility in formal settings but favor stylistic flair in casual contexts. State-of-the-art methods, such as \cite{liu2024unifying, guan2022personalized}, employ attention mechanisms to adaptively weight compatibility and personal-preference signals, however, they still rely on manually defined fusion rules rather than data-driven integration. Another recent work CP-TransMatch~\cite{transmatch} uses a novel TransE expansion model to align user product-pair triplet interactions, yet it lacks explicit mechanisms to preserve contextual-aware substructures.    

\textit{Existing models often treat product-compatibility and personal-preference as either independent factors or decoupled correlated dimensions, failing to account for their overlap and dynamic contextual interactions. Moreover, most methods implicitly assume that all user-product interactions reflect true preferences, disregarding the noisy nature of such interaction data -- often arising from accidental clicks or mismatched outfits. }

\begin{figure*}[t]
  \centering
  \includegraphics[width=\linewidth]{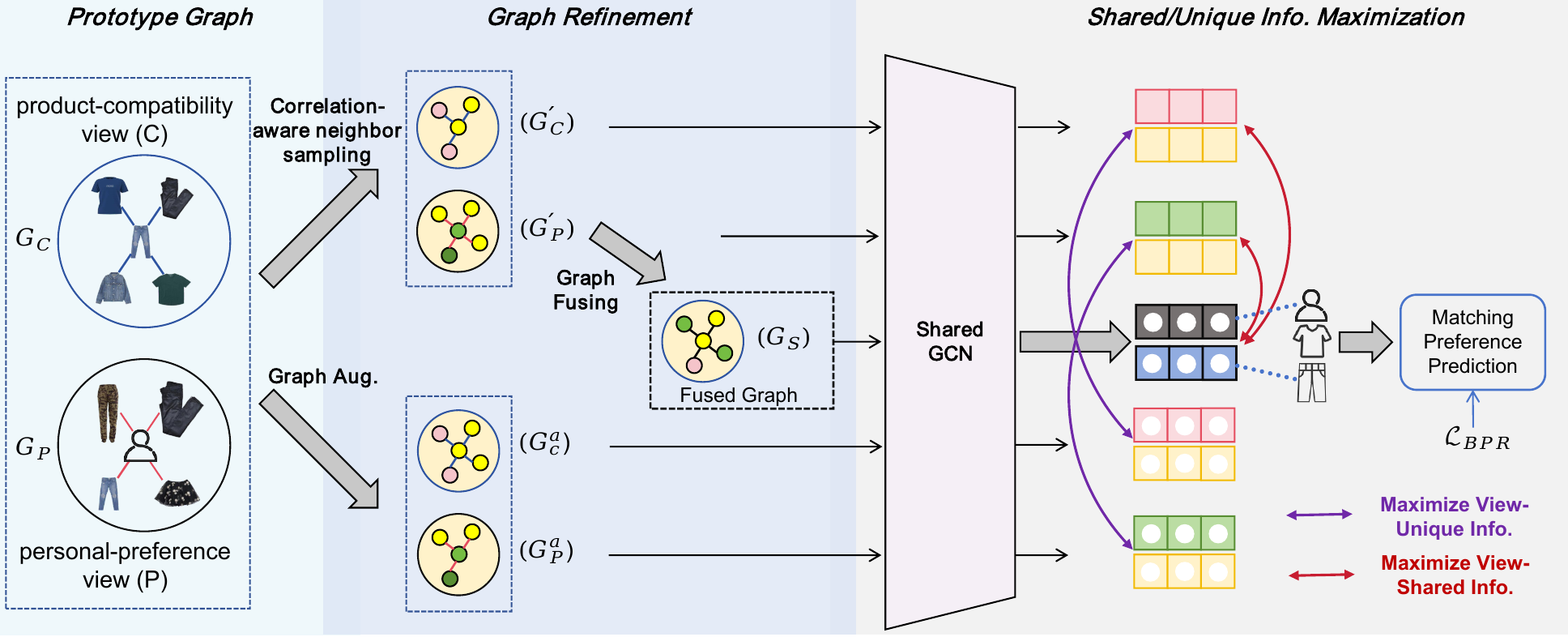}
  \caption{ The overall framework of the proposed PCGNet. PCGNet first builds prototype graphs for personal-preference and product- compatibility, then generates refined graphs and fused graphs through graph learning. Subsequently, it maximizes both shared and view-specific information after shared GCN encoder, and the learned node representation are used for personalized fashion matching preference prediction. }
  \label{fig: main}
  % \vspace{-8px}
\end{figure*}
\subsection{Unsupervised Graph Structure Learning } 
Graph structure learning plays a pivotal role in recommendation systems, particularly when handling noisy and sparse data. Traditional GCN-based recommenders, such as NGCF \cite{NGCF2019}, propagate information through raw interaction graphs, which inevitably amplifies noise \cite{jin2020graph, liao2024hypergraph}. To address this, recent studies have explored \textit{robust graph construction}. \citet{liu2022rgcf} refined adjacency matrices by pruning low-confidence edges, but their static pruning strategy fails to adapt to diverse user scenarios. \citet{jin2020graph} proposed adversarial training to defend against edge perturbations, yet their emphasis on security rather than task relevance limits its applicability to fashion recommendations.  
A promising direction involves \textit{subgraph learning combined with mutual information maximization}. Inspired by Deep Graph Infomax (DGI) \cite{velivckovic2018deep}, \citet{graphsurvey} leveraged global-local mutual information for better graph representations. However, these methods primarily target homogeneous graphs and struggle with heterogeneous fashion graphs, which contain both user-product and product-product interactions. \citet{fan2019graph} introduced heterogeneous GNNs to model multi-typed nodes/edges, yet their approach does not explicitly address noise suppression. 

\textit{In brief, most graph-based learning methods rely on a `one-size-fits-all' noise removal strategy, failing to distinguish between compatibility-critical and preference-critical edges. Existing mutual information frameworks primarily optimize single-view representations, rather than facilitating cross-view alignment between compatibility and preference graphs. }

\section{Methodology}
Fig.~\ref{fig: main} shows the main structure of PCGNet, which contains three core components. We first construct two graph views that represent personal preferences $G_P$ and product compatibility $G_C$, respectively.
Next, in the graph structure refinement module, we extract correlation aware sub-graphs from the original graph for each view to generate more reliable graph structure ($G_P^{'}$, $G_C^{'}$). Moreover, a contextual aware graph augmentation branch is also leveraged to generate augmented graphs ($G_P^{a}$, $G_C^{a}$) for the two views. These augmented graphs are designed to solve the problem of insufficient learning of latent features caused by interaction sparsity, while removing irrelevant noise that are not related to our product-compatibility/personal-preference targets in the original graph. We then fuse the graphs from both views into a comprehensive graph representation $G_{S}$ for generating the final prediction through the graph fusion module. This fusion graph aims to integrate the information related to the recommendation goal from all views to generate a more comprehensive recommendation result ($\zeta_r^{u,g}$). Additionally, information entropy is particularly adopted to handle non-redundancy in multi-view graph data. Different views may contain unique task-relevant information, which information entropy can effectively capture by maximizing mutual information \cite{2024beyond}. In our training process, we not only utilize the traditional BPR loss but also incorporate mutual information maximization based on information entropy for both cross-view and specific-view patterns.
% Additionally, since information entropy can handle non-redundancy in multi-view graph data, and different views may contain unique task-relevant information. Information entropy can capture this unique information by maximizing Mutual Information \cite{2024beyond}. Besides the traditional BPR loss, we also employed the information entropy based mutual information maximization for both cross view and unique views in the training process.
% During training process, we adopt 
\subsection{Problem Formulation}
We formulate the personalized fashion matching task as follows. Let $\mathcal{U} = \{u_1, ..., u_{|\mathcal{U}|}\}$ denote the user set, $\mathcal{G} = \{g_1, ..., g_{|\mathcal{G}|}\}$ the set of given fashion items (query products, e.g., top clothing), and $\mathcal{R} = \{r_1, ..., r_{|\mathcal{R}|}\}$ the set of recommendable matching items (e.g., bottom clothing). Each valid outfit combination $\langle g, r \rangle$ with $g \in \mathcal{G}, r \in \mathcal{R}$ receives implicit feedback through user interactions $\mathcal{O} = \{(u, g, r) | u \in \mathcal{U}\}$, where positive triplets indicate observed engagements (clicks or purchases).
Each product $i \in I = \mathcal{G} \cup \mathcal{R}$ is represented by: 1) visual features $\mathbf{v}_i \in \mathbb{R}^{d_v}$ from product images, and 2) textual features $\mathbf{t}_i \in \mathbb{R}^{d_t}$ from product descriptions. 
% Users and items also have latent embeddings $\mathbf{e}_u \in \mathbb{R}^d$ and $\mathbf{e}_r \in \mathbb{R}^d$ for collaborative filtering, following matrix factorization principles~\cite{BPR,vbpr}.

\textbf{Personalized Fashion Matching Task.} Our goal is to learn a recommendation model $\mathcal{F}$ that recommend a fashion item $r$ to match with a given
garment $g$ for a specific user $u$, and
predict the personalized compatibility score for triplet $\langle u, g, r \rangle$:
\begin{equation}
    \zeta_{^{u,g}_r} = \mathcal{F}(u, g, r | \Theta)
    \label{eq:overall_obj}
\end{equation}
where $\Theta$ denotes model parameters. The score $\zeta_{^{u,g}_r}$ should reflect both the user's personal preference and the product compatibility between $g$ and $r$. Following established settings in fashion recommendation~\cite{GPBPR, PCE, CP}, we focus on two key fashion categories: \textit{top garments} (shirts, blouses, etc.) and \textit{bottom garments} (pants, skirts, etc.), where $\mathcal{G}$ and $\mathcal{R}$ represent complementary categories (e.g., recommending bottoms for top queries). The model should score positive triplets higher than negative/unobserved ones. 

% \subsection{PCGNet}
\subsection{Graph Construction and Node Embedding}
In our approach, we construct heterogeneous graphs based on two complementary views. In most previous studies \cite{guan2022personalized, yang2023heterogeneous}, representations were learned from complicated relations between users and products within a single graph, which may easily lead to incomplete learning of interactions and inter-dependencies among users or items. On the other hand, even if two fashion products are compatible, they might not be a good recommendation if they do not align well with the user's preferences \cite{liu2024unifying}. %, the unique importance of product compatibility/personal preference are overlooked in previous studies. Hence 
Instead, we formulate heterogeneous graphs through two complementary views. \textit{Product-Compatibility Graph} ($G_C=\{A_C, E_V\}$) captures intrinsic matching relationships between fashion products. The adjacency matrix $A_C \in {0,1}^{V \times V}$ is derived from co-occurrence statistics in valid outfits, where edges connect compatible tops and bottoms (e.g., a striped shirt paired with denim jeans). Nodes inherit unified features from $E_V$, which aggregates attributes of users and items ($V=\mathcal{U}\cup I$).
The complementary \textit{Personal Preference Graph} ($G_P=\{A_P, E_V\}$)
captures user-item interaction behaviors through the adjacency matrix $A_P$. To more comprehensively model the differentiated preferences of users for tops and bottoms, we not only retain the user-bottom interaction edges to reflect preferences for recommendable matching items but also introduce user-top interaction edges to explicitly characterize users' selection for given items. The edge weights are quantified by the frequency of selection to represent the strength of preference. This design enables $ G_P $ to capture the overall dressing preferences of users for both tops and bottoms, rather than relying solely on historical behaviors of matching items. Both graphs share identical node sets while maintaining separated edge semantics. This special design ensures that subsequent learning processes can disentangle compatibility signals from personalization patterns.

% models personalized taste via user-product interactions, and $A_P$ defines adjacency matrix where edges connect users to preferred items. Both graphs share the unified node set, preserving cross-view structural alignment while differentiating edge semantics.
% we draw inspiration from constructing both homogeneous (i.e. item-item graph) and heterogeneous (i.e. user-item) relation graphs \cite{liu2024unifying}, to model these relationships comprehensively from both item-compatibility space (noted as C) and personal-preference space (noted as P). The \textit{item-compatibility homogeneous graph ($G_{ii}$)} is designed to model the top-bottom compatibility relations, while the \textit{personal-preference heterogeneous graph $G_{ui}$} is designed to model the user's personalized preference towards the target product to be recommended. 

\begin{figure}[t]
  \centering
  \setlength{\abovecaptionskip}{0.2cm}
  \includegraphics[width=\linewidth]{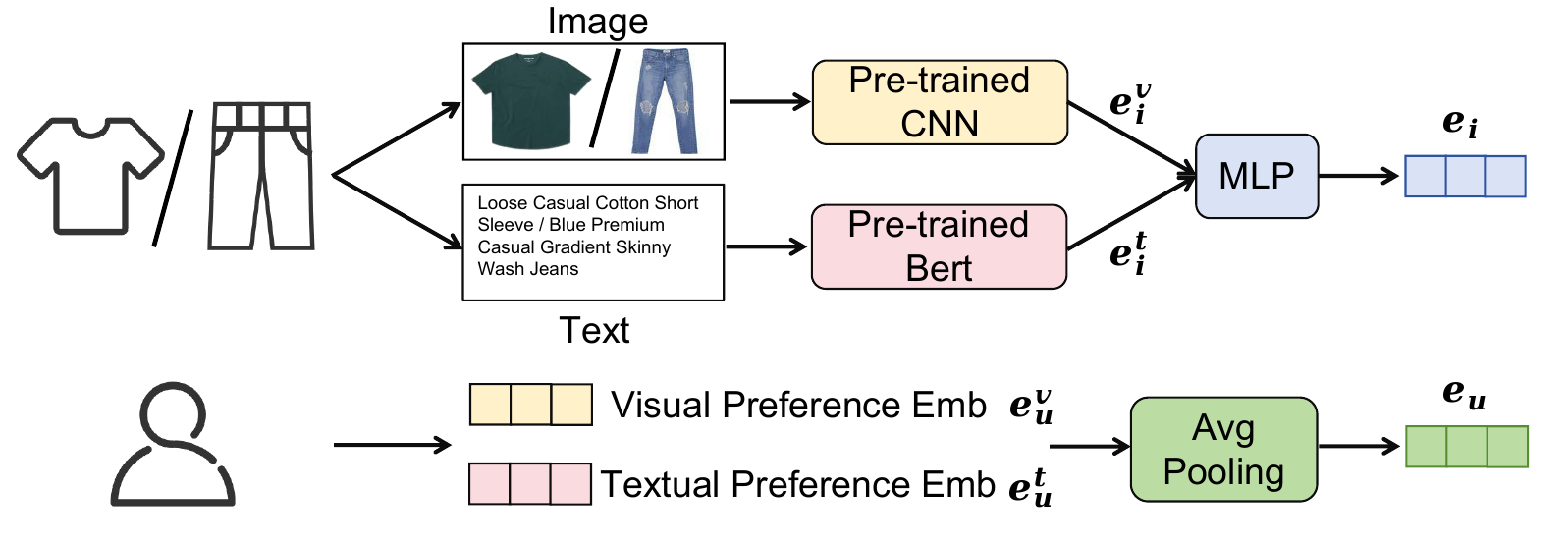}
  \caption{Item matching and personal preference graph initial node embedding.}
  \label{fig: node}
  \vspace{-5px}
\end{figure}

For the generation of initial node embeddings, given the significant differences in entity features between users and items (tops/bottoms), we adopt a categorical encoding strategy. The initial representations of item nodes are derived from visual features and textual category attributes, while user nodes are initialized with learnable latent factors, separately learning users' preferences for product visual and textual attributes.

% For the detailed node initial embedding of each item(top/bottom) and user, we first derive the initial node-level representations in the item matching compatibility graph and personal preference graph.
% As the users and items are two types of entities, and
% the node contents differ remarkably, we learn their embeddings separately as shown in Fig. \ref{fig: node}. 

\textbf{Item Entity Embedding.} Each item entity consists of an image and a text description, for any given item (top or bottom). We employ ResNet to extract its visual features ($\mathbf{v}_i$) and a pre-trained BERT model to derive its textual features ($\mathbf{t}_i)$). Subsequently, we concatenate the visual and textual features of each item $i$ to form its final embedding. This embedding is then projected onto a lower-dimensional space using a learnable fully-connected layer. Mathematically, this process can be expressed as:
% \begin{equation}
% \left\{
% \begin{aligned}
% &\mathbf{e}_i^v = \operatorname{ResNet}(\mathbf{v}_i),\\
% &\mathbf{e}_i^t = \operatorname{BERT}(\mathbf{t}_i),\\
% &\mathbf{e}_i^{(P/C)} = \operatorname{{MLP}_{P/C}}([\mathbf{e}_i^v, \mathbf{e}_i^t]),\\
% \end{aligned}
% \right.
% \label{eq: item_node}
% \end{equation}
\begin{equation}
\left\{
\begin{aligned}
&\mathbf{e}_i^v = \operatorname{ResNet}(\mathbf{v}_i),\\
&\mathbf{e}_i^t = \operatorname{BERT}(\mathbf{t}_i),\\
&\mathbf{e}_i = \operatorname{MLP}([\mathbf{e}_i^v, \mathbf{e}_i^t]),\\
\end{aligned}
\right.
\label{eq: item_node}
\end{equation}
where $[,]$ refers to the concatenation operation, and $\mathbf{e}_i \in \mathbb{R}^D$ is the final entity embedding of item $i$; $D$ is the dimension of the embedding.
% The $P$ in the superscript indicates that it is exclusively to the personal preference space; while $C$ indicates that it is exclusively to the product compatibility space, so as to distinguish the latent embedding of the items on different graphs.
 
\textbf{User Entity Embedding.} Instead of using traditional one-hot embeddings, we use a learnable visual latent preference factor ($\mathbf{e}_u^v$) and a textual latent preference factor ($\mathbf{e}_u^t$) to capture each user's potential visual and textual preferences for the fashion product \cite{VBPR, GPBPR}. Additionally, a cold-start user's embedding can also be derived. The user initial embedding ($\mathbf{e}_u$) for a user $u$ is then constructed as follows:
\begin{equation}
    \mathbf{e}_u = AvgPool(\mathbf{e}_u^v, \mathbf{e}_u^t),
\end{equation}
where $\mathbf{e}_u \in \mathbb{R}^D$, and $D$ is the dimension of the user embedding, which is the same as that of item embedding.

\subsection{Graph Refinement}

\subsubsection{Correlation-aware Neighbor Sampling.} 
The \textit{correlation-aware neighbor sampling} strategy \cite{liu2024unifying} is to acknowledge the contribution of neighboring nodes in feature aggregation towards the target node. Taking a user $u$ in the personal-preference prototype graph $G_{P}$ as an example, to highlight the importance of different item nodes, we first calculate the correlation sub-matrix $\Phi_m^P\in \mathbb{R}^{n \times 1}$ between the target user $u$ and its neighboring product nodes:
\begin{equation}
\Phi_m^P = \operatorname{Softmax}(\operatorname{sim}(e_u, E_{In})),
\end{equation}
where $E_{In} \in \mathbb{R}^{n \times D}$ is the feature matrix of the product nodes, where $n$ and $D$ represent the number of products and the corresponding vector dimension, respectively. The function $\operatorname{sim}$ computes the cosine product between the representation $e_u \in \mathbb{R}^D$ of the target user node and each of its neighboring product nodes. The closer the value of each row in $\Phi_m^P$ (approaching to 1), the higher the correlation between the user and the product. This strategy is adopted to help select relevant nodes and mitigate the impact of noisy nodes during feature aggregation. The correlation-aware neighbors sampling is as follows:
\begin{equation}
\left\{
\begin{aligned}
& n_k = f_c(u_d \cdot p | \Phi_m^P),\\
& p = \left(\frac{1}{2}\right)^j \quad (j = 1, 2, 3 \ldots),\\
\end{aligned}
\right.
\label{eq: node_sampling}
\end{equation}
where $n_k$ is the sampled neighbor item nodes, $u_d$ is the degree of the target user node $u$ and $p$ is the sampling probability. $j$ represents hop, and $f_c$ is the ceiling function. The top $k$ correlative item neighbors are sampled based on the sub-matrix $\Phi_m^P$ to participate in the subsequent node feature aggregations. 
% After performing the the same sampling process, we can get the other item-compatibility correlation aware refined graph $G_C'$.

Then to illustrate the node feature aggregation process, we still take the $G_P$ as an example. Specifically, the latent representation of a target user $e_u^{P(l)}$ at the $l-th$ graph neural layer at personal preference view (P) after feature aggregation can be expressed as:
\begin{equation}
\left\{
\begin{aligned}
& e_u^{P(l)} = \sigma\left(W \cdot \left(e_u^{P(l-1)} \oplus \text{Agg}_{P}\{e_m^{P(l-1)}, m \in n_k^P\}\right) + b\right),\\
& \text{Agg}_{P}\{e_m^{P(l-1)}, m \in n_k^P\} = \sum_{m=1}^{|n_k^P|} \alpha_m^{l-1} e_m^{P(l-1)};\\
\end{aligned}
\right.
\label{eq: node_feature_aggre}
\end{equation}
where $W$ and $b$ represent the weight and bias, respectively. $\sigma$ denotes the sigmoid activation function. The symbol $\oplus$ represents vector concatenation operation. $n_k^P$ represents all the neighbors of the target user $u$, which are sampled by Eq. (\ref{eq: node_sampling}) for personal preference (P) modeling. An attention mechanism within the feature aggregation process is adopted, we represent the relative importance of each neighboring to user $u$ as:
\begin{equation}
\left\{
\begin{aligned}
&\alpha^{(l)*}_m = e_u^{P(l)} \cdot e_m^{P(l)} \\
&\alpha_m = \frac{\exp(\alpha^{(l)*}_m)}{\sum_{m \in N_k^P} \exp(\alpha^{(l)*}_m)}\\
\end{aligned}
\right.
\end{equation}
Then, we can use the same neighbor sampling strategy (\ref{eq: node_sampling}) and node feature aggregation process (\ref{eq: node_feature_aggre}) to obtain the updated refined graphs $G_P' =\{A_P',E_P'\}$ and $G_C' =\{A_C',E_C'\}$, where the updated $\Phi_m^P \in A_P'$, and $e_u^P \in E_P'$ .
% , where $\Phi_m^P \in A_P'$, and $e_u^{P(l)} \in E_P'$.

\subsubsection{Learnable Graph Augmentation}
We adopt the learnable generative augmentation strategy \cite{2024beyond} to specifically handle two views (P/C), aiming to improve the graph structure by generating augmented graphs that capture product-compatibility or personal-preference specific information while eliminating noise. 
The key idea is to use a learnable graph augmentation generator to create personalized edge weights for existing edges in each view, ensuring that the augmented graphs retain only objective-relevant information. 
For each of the P and C views, we generate an augmented graph $G_P^{a}$ and $G_C^{a}$, respectively. 
% The augmented graphs are designed to maximize the unique information with the refined graphs $G_P^{'}$ and $G_C^{'}$. 
Taking P view as an example, for each user-product edge $a_{ui}$ in the parent view, we compute its edge weight ($w_{ui}^P$) by:
% \begin{equation}
% w_{ui}^P= \frac{e_u \cdot e_i}{\|e_u\| \|\mathbf{e_i\|},
% \label{eq: topk_sim_w}
% \end{equation}

% using a Multilayer Perceptron (MLP) based on the node features:
\begin{equation}
\left\{
\begin{aligned}
&\theta_{ui}^{P} = \text{MLP}([e_u, e_i]), \\
&\omega_{ui}^{P} = \text{Sigmoid}\left(\frac{\log \delta - \log(1 - \delta) + \theta_{ui}^{P}}{\tau}\right),\\
\end{aligned}
\right.
\end{equation}
$\delta \sim Uniform(0,1)$ is a sampled Gumbel \cite{maddison2016concrete, jang2016categorical} random variate, and $\tau$ is the temperature that controls the sharpness of the distribution. Here, by approaching $\tau$ to zero, we make $w_{ui}^P$ to become a binary distribution \cite{2024beyond}.
The reconstructed graph $G_P^a =\{A_P^a,E_P^a\}$ and $G_C^a =\{A_C^a,E_C^a\} $ are then obtained using the MLP-based decoder of the augmented views.

\subsubsection{Graph Fusion}
The \textit{Graph Fusion} process is to generate a fused graph $G_S$ that jointly leverage personal-preference and product-compatibility information from the refined graphs $G_P^{'}$ and $G_C^{'}$.
This fused $G_S$ should be more informative compared to individual views, making it suitable for our personalized fashion matching recommendation.
The fused graph learner is defined as:
\begin{equation}
E_{S} = \sigma\left([E_C', E_P']\odot W_1\right) \odot W_2.
\label{eq: fusing}
\end{equation}
where $\sigma$ is the ReLU activation function, and $W_1$, $W_2$ are learnable weight matrices. The node features from both views ($E_C', E_P'$) are concatenated to form a combined feature matrix, and we weigh the importance of features from different views.

\subsection{Shared \& Specific Information Maximization}
\subsubsection{Shared GCN and Matching Preference Prediction}
To ensure consistency and comparability of the node representations, we utilize a shared encoder across different views. The shared encoder allows for the integration of features from different views in a unified manner. This integrated learning helps in capturing both shared and unique information across views. Here the shared graph encoder is implemented using a typical Graph Convolutional Network (GCN) \cite{kipf2016semi}, which is proven well-suited for learning node representations in graph-structured data \cite{2024beyond}. To obtain the node representations, suppose we have a given graph $G_j=\{A_j, E_j\}$ from the $j\-th$ view, we process:
\begin{equation}
Z_j = \text{GCN}(A_j, E_j),
\label{eq: GCN}
\end{equation}
where the node representation $Z_j \in \mathbb{R}^{V\times D}$, and the parameters of the GCN layers are shared across all views. Then the final matching preference score \( \zeta_{^{u,g}_r} \) for matching product $r$ to be recommended for given item $g$ for 
each user $u$ is then computed as:
\begin{equation}
  \zeta_{^{u,g}_r} = MLP[Z_u^s, Z_g^s, Z_r^s],
\end{equation}
where $[,,]$ represent concatenation and $Z_u^s, Z_g^s, Z_r^s$ are extracted from the node embedding $Z_{S}$ obtained from Eq. (\ref{eq: GCN}).

\subsubsection{Joint Optimization}
The model is trained end-to-end by minimizing a widely used Bayesian Personalized Ranking (BPR) over observed triplets $(u, g, r)$. At the same time, we maximize the mutual information not only from view-shared but also view-specific information. The loss terms for minimization can be expressed by $\mathcal{L}_\cap$ for view-shared perspective, and $\mathcal{L}_\varnothing$ for view-unique perspective. The jointly optimization process can be formulated as:

\begin{equation}
\left\{
\begin{aligned}
&\mathcal{L}_{BPR} = \sum_{\mathcal{D}}\left[-\ln \left(\sigma\left(\zeta_{r_+}^{u,g}-\zeta_{r_-}^{u,g}\right)\right)\right]+\frac{\lambda}{2}\left\|\Theta_F\right\|^2,\\
& \mathcal{L}_\cap = -\left[ \mathcal{I}\left(\mathbf{Z}_S, \mathbf{Z}_C'\right)+ \mathcal{I}\left(\mathbf{Z}_S, \mathbf{Z}_P'\right)\right],\\
& \mathcal{L}_\varnothing = -\left[\mathcal{I}\left(\mathbf{Z}_C^a, \mathbf{Z}_C'\right)+ \mathcal{I}\left(\mathbf{Z}_P^a, \mathbf{Z}_P'\right)\right],\\
& \mathcal{L} = \mathcal{L}_{BPR} + \mu \cdot(\mathcal{L}_\cap + \mathcal{L}_\varnothing)
\end{aligned}
\right.
\label{eq: loss_all}
\end{equation}

where $\mu$ balances the two objectives. BPR aims to maximize the margin between positive and negative samples, ensuring higher rankings for positive candidates. Given a pair of positive and negative samples, their preference scores, computed using Eq. (\ref{eq:overall_obj}), are denoted as $\zeta_{r_+}^{u,g}$ and $\zeta_{r_-}^{u,g}$, respectively. $\mathcal{D}$ is the training set.
$\mathcal{I}(\cdot)$ denotes mutual information \cite{2024beyond},
% estimated via \textbf{information entropy} \cite{2024beyond}.
% Mutual information is 
% a measure of the amount of shared information between two random variables,
which is widely used to measure the degree of interdependence between the two variables. Taking $I(Z_S; Z_C')$ as an example:
% \begin{equation}
% I(G_1; G_2) = \sum_{g_1 \in G_1} \sum_{g_2 \in G_2} p(g_1, g_2) \log \frac{p(g_1, g_2)}{p(g_1) p(g_2)}
% \label{eq: mutual_loss}
% \end{equation}
\begin{equation}
I(Z_S; Z_C') = \sum_{z_s \in Z_S} \sum_{z_c' \in Z_C'} p(z_s, z_c') \log \frac{p(z_s, z_c')}{p(z_s) p(z_c')},
\label{eq: mutual_loss}
\end{equation}
where $p(z_s, z_c')$ denotes the joint distribution of node representations
from the shared view $S$ and refined product-compatibility view $C'$, while $p(z_s)$ and $p(z_c')$ denotes the marginal distribution, respectively. $z_s$ and $z_c'$ are the representations of the same node in views $S$ and $C'$. For example, $I(Z_S; Z_C')$ measures how much overlap between shared preferences and compatibility rules.
Due to the complexity of graph structured data, and we focus on node-level task, here we only use node embeddings for mutual information calculation. Additionally, node embeddings provide a compact and meaningful representation that can capture both local and global graph structures, which assume containing sufficient task relevant information \cite{liu2024towards}.
 
% Here we can compute the mutual information between two graphs through node embedding：
% \begin{equation}
% I(E_1; E_2) = \sum_{e_1 \in E_1} \sum_{e_2 \in E_2} p(e_1, e_2) \log \frac{p(e_1, e_2)}{p(e_1) p(e_2)}
% \end{equation}

\section{Experiments}
To evaluate the effectiveness of PCGNet, we conducted extensive experiments on two benchmark datasets, giving answers to the following research questions (RQs):

\begin{itemize}
    \item \textbf{RQ1:} Can the proposed PCGNet outperform previous personalized fashion matching methods?
    \item \textbf{RQ2:} Can the contextual aware neighbor sampling strategy and shared/unique information maximization modules help improve item matching accuracy?
    \item \textbf{RQ3:} How do specific settings in the method affect overall performance?
\end{itemize}

\subsection{Datasets}
We utilized two benchmark fashion datasets with personalized item matching data for our experiments: the IQON3000 \cite{GPBPR} and Polyvore-U-519 \cite{binary} datasets. These datasets are widely used for comparative studies in fashion matching recommendation research. 
% The statistics of the datasets are summarized in Table \ref{dataset_statistics}. 

\noindent The \textbf{IQON3000} dataset was collected from a Japanese website IQON, which consists of 308,747 fashion outfits created by 3,568 users, comprising a total of 672,335 fashion items. 
With a focus on personalized fashion matching recommendation, we therefore adopted the experiment setup as that for GP-BPR study \cite{GPBPR} and filtered subset of IQON3000 for experiment.

\noindent The \textbf{Polyvore-U} dataset contains user-composed outfits, specifically adopting the Polyvore-519 version, which includes a larger number of users and user-top-bottom triplets. Focusing on fashion matching, we filtered a subset containing only pairs of tops and bottoms. The statistics of dataset after pre-processing are summarized in Table \ref{dataset_statistics}. 

The PCGNet model recommends matching clothing by learning multi-modal representations, covering visual and textual features. For IQON3000, 2048-D visual features from ResNet50 and textual features from pre-trained BERT were used. For Polyvore-U-519, 2400-D textual features from AlexNet and 2048-D visual features from ResNet152 were utilized. For each user, based on the principle of anonymity and privacy protection, there is no personally identifiable information, only user IDs were used; the recommendation task under such configuration is more challenging.

\begin{table}[!t]
\caption{Statistics of datasets used for experimental analysis}
\centering
\renewcommand{\arraystretch}{1.1}
\setlength{\tabcolsep}{8pt}
\begin{tabular}{lrr}
\toprule
% \multirow{2}{*}{C} & \multicolumn{2}{c}{Dataset} \\
% \cmidrule(lr){2-3}
Dataset & Polyvore-519 & IQON3000 \\
\midrule
\textbf{User Statistics} &519 & 3,236  \\
% \quad Number of users & 519 & 3,236 \\
\addlinespace
\textbf{Garment Statistics} & & \\
\quad Top garments & 16,887 & 99,655 \\
\quad Bottom garments & 16,220 & 43,086 \\
\quad \textbf{Total garments} & \textbf{33,107} & \textbf{142,737} \\
\addlinespace
\textbf{Sample Statistics} & & \\
\quad Training samples & 39,133 & 170,601 \\
\quad Validation samples & 5,110 & 23,095 \\
\quad Testing samples & 5,197 & 23,095 \\
\quad \textbf{Total samples} & \textbf{49,440} & \textbf{216,791} \\
\bottomrule
\end{tabular}
\label{dataset_statistics}
\end{table}

\subsection{Experimental Setting}
%\subsubsection{Baselines}
\subsubsection{State-of-the-Art Methods for Comparative Study}
To demonstrate the effectiveness of PCGNet, 
we compared our method with a set of personalized fashion matching recommendation SOTA methods, including 
\textbf{MF-BPR} ~\cite{BPR}, \textbf{VBPR} ~\cite{VBPR}, \textbf{T-BPR}~\cite{GPBPR}, \textbf{VT-BPR}~\cite{GPBPR}, \textbf{GP-BPR}~\cite{GPBPR}, \textbf{PCE-Net}~\cite{PCE}, \textbf{DGSR} \cite{DGSR}, \textbf{TransRec+V} \cite{TransRec}, \textbf{CP-TransMatch}~\cite{CP}, and \textbf{TryonCM2}(2025) \cite{TryonCM2}.
The main characteristics of these baselines are summarized in Table \ref{tab:method_features}.
\begin{table}[!t]
\caption{Characteristics matrix of baseline methods}
\centering
\small
\begin{tabular}{lccccccc}
\toprule
 & \rotatebox{60}{Visual} & \rotatebox{60}{Textual}  & \rotatebox{60}{Graph} & \rotatebox{60}{Attention} & \rotatebox{60}{P} & \rotatebox{60}{C} \\
\midrule
BPR-MF~\cite{BPR}  & &  & & &\checkmark & \\
V-BPR~\cite{VBPR}  & \checkmark & &  & &\checkmark & \\
T-BPR~\cite{GPBPR} & & \checkmark &  & &\checkmark &\\
VT-BPR~\cite{GPBPR}  & \checkmark & \checkmark & & &\checkmark &\\
GP-BPR~\cite{GPBPR}  & \checkmark & \checkmark & & &\checkmark &\checkmark\\
PCE-NET~\cite{PCE}  & \checkmark & \checkmark &  & \checkmark &\checkmark &\checkmark\\
DGSR~\cite{PCE}  & \checkmark & \checkmark &  \checkmark & &\checkmark &\checkmark\\
TransRec+V~\cite{CP}  &\checkmark  & & \checkmark & &\checkmark &\\
CP\_TransMatch~\cite{CP}  &\checkmark  & & \checkmark & &\checkmark &\checkmark\\
% NiPC-BPR~\cite{liao2023recommendation} & \checkmark & \checkmark & &  &\checkmark &\checkmark\\
TryonCM2~\cite{TryonCM2}  & \checkmark &  & & \checkmark &\checkmark &\checkmark\\
\bottomrule
\end{tabular}
\label{tab:method_features}
\end{table}

\begin{itemize}
    \item \textbf{MF-BPR} ~\cite{BPR}: Applies matrix factorization on the user-item interaction matrix to capture user preferences and optimize the model using Bayesian Personalized Ranking.
    \item \textbf{VBPR} ~\cite{VBPR}: An extension of MF-BPR that includes additional visual preference modeling by uncovering visual user-item interaction patterns.
    \item \textbf{T-BPR}~\cite{GPBPR}: Similar to V-BPR but uses textual features.
    \item \textbf{VT-BPR}~\cite{GPBPR}: Combines V-BPR and T-BPR for comprehensive user preference modeling.
    \item \textbf{GP-BPR}~\cite{GPBPR}: A personalized item matching method that decomposes the problem into modeling personal preferences and product compatibility, treating these components separately.
    % \item \textbf{PAI-BPR:} A follow-up method based on GP-BPR that leverages fine-grained visual features based on item attributes to boost performance.
    \item \textbf{PCE-Net}~\cite{PCE}: Uses attention-based modeling for product matching and personal preference.
    \item \textbf{DGSR} \cite{DGSR}:A method for sequential fashion recommendation that models third-order interaction data among users and fashion items.
    \item \textbf{TransRec} \cite{TransRec}: A translation-based method for sequential recommendation, learning the latent transition space where item transitions made by different users can be measured with embeddings.
    \item \textbf{TransRec+V} \cite{TransRec}: An extended version of TransRec that models additional visual transition patterns.
    \item \textbf{CP-TransMatch}~\cite{CP} captures third-order user-item-item interactions. It constructs a multi-relational graph where items are entities and users are relations, and enhances the translation-based matching with context and path modules. 
    % \item \textbf{MG-PFCM:} The state-of-the-art personalized fashion matching approach that aggregates user and item pairs to obtain overall representations.
    \item \textbf{TryonCM2} \cite{TryonCM2}:  integrates both item interactions and synthesized try-on appearance using bidirectional LSTMs to enhance outfit compatibility assessment.
\end{itemize}

\subsubsection{Implementation Details and Evaluation Metrics}
We applied the Adam \cite{ADAM} optimizer to train all compared models using the PyTorch framework, tuning training parameters independently for each method on each dataset.  A grid search strategy was employed to optimize training parameters. The mini-batch size was varied within [64, 128, 256, 512], weight decay within [0.001 to 0.0000001], hidden dimensions within [256, 512], and learning rate within [0.001, 0.0001, 0.00001]. During training, the online negative sampling was applied. For a fair comparison, the hidden dimension was set to 32 for all methods. For model comparison with DGSR \cite{DGSR} and TransRec+V \cite{TransRec}, we report the performance as presented in the original paper \cite{CP} and compare to our method based on exactly the same dataset that they were tested on.

\begin{table*}[!t]
\caption{Performance comparison. \textbf{Bold} data indicate the best results, and \underline{\textit{underlined}} ones are the second best results.}
% \vspace{-8px}
\begin{center}
\setlength{\tabcolsep}{3mm}
\begin{tabular}{l|cccc|cccc}
\toprule
\multirow{2}{*}{Method}&\multicolumn{4}{c|}{Polyvore-U-519} & \multicolumn{4}{c}{IQON3000} \\
\cmidrule(lr){2-5} \cmidrule(lr){6-9}
& AUC$\uparrow$ & HR@10$\uparrow$& NDCG@10$\uparrow$& MRR@10$\uparrow$& AUC$\uparrow$& HR@10$\uparrow$& NDCG@10$\uparrow$& MRR@10$\uparrow$ \\
\midrule
MF-BPR \cite{BPR} & 0.7639 & 0.6916 & 0.5434 & 0.4973 & 0.8309 & 0.7024 & 0.5243 & 0.4687 \\
T-BPR \cite{GPBPR}& 0.7839 & 0.6502 & 0.4868 & 0.4359 & 0.8316 & 0.6361 & 0.4301 & 0.3666 \\
V-BPR \cite{VBPR}& 0.8074 & 0.7141 & \underline{\textit{0.5848}} & \underline{\textit{0.5443}} & 0.8356 & 0.6941 & 0.5230 & 0.4694 \\
VT-BPR \cite{GPBPR}& 0.8105 & 0.6846 & 0.5266 & 0.4772 & 0.8384 & 0.7003 & 0.5134 & 0.4551 \\
DGSR \cite{DGSR}& 0.8320 &0.6394 &-- &-- & 0.8742 & 0.6645 &-- &--\\
GP-BPR \cite{GPBPR}& 0.8232 & 0.7121 & 0.5716 & 0.5277 & 0.8569 & 0.7396 & 0.5849 & 0.5366 \\
% TransRec &0.8229 & 0.6353 &-- &-- & 0.8588 & 0.6561 &-- & -- \\
TransRec+V \cite{TransRec} & 0.8284 & 0.6414 &-- &-- &0.8694 &0.6575 &-- &--\\
PCE-NET \cite{PCE}& 0.8235 & 0.4208 & 0.2380 & 0.1825 & 0.8341 & 0.6399 & 0.4110 & 0.3399 \\
CP-TransMatch \cite{CP} & \underline{\textit{0.9001}} & \underline{\textit{0.8573}} & 0.5472 & 0.5144 & \underline{\textit{0.8842}} & \underline{\textit{0.8789}} & \underline{\textit{0.6453}} & \underline{\textit{0.5430}} \\
TryonCM2(2025) \cite{TryonCM2} & 0.8730 & 0.4640 &  0.3310 & 0.3030& --&--&--&--\\
\midrule
\textbf{PCGNet} & \textbf{0.9667} & \textbf{0.8814} & \textbf{0.7117} & \textbf{0.6704} & \textbf{0.9725} & \textbf{0.9129} & \textbf{0.7386} & \textbf{0.6802} \\
% \% improvement & \textbf{7.40\%} &\textbf{2.81\%} & \textbf{21.70\%} & \textbf{23.17\%} &
% \textbf{9.97\%} &\textbf{3.87\%} & \textbf{14.46\%} & \textbf{25.27\%} \\
\bottomrule
\end{tabular}
\end{center}
\label{tab: overall}
% \vspace{-2px}
\end{table*}

\textit{Evaluation Protocols.} Each training triplet consists of a positive sample, with a negative sample generated by randomly selecting another bottom product from the product set that has not been interacted with the corresponding user before. For evaluation metrics, we employed four evaluation metrics to assess the PCGNet model's effectiveness: AUC, HR@$k$, NDCG@$k$, and MRR@$k$. 
\begin{itemize}
\item The Area Under the Curve (AUC) evaluates the model's ability to correctly identify positive matches over negatives across the testing/validation set. The best model performance is determined by the AUC on the testing set at the epoch where the validation set achieves its peak performance.

\item The Hits Ratio (HR@$k$) measures the proportion of relevant products within the top recommendations, reflecting user satisfaction. For the HR test, we randomly sample $k = 99$ negatives and rank the positive and negative products and calculate the top $k$ hit rate HR@$k$. 

\item Normalized Discounted Cumulative Gain (NDCG@$k$) considers both the relevance and position of products in the ranking list.

\item Mean Reciprocal Rank (MRR@$k$) evaluates ranking performance based on the position of the first relevant product. 
\end{itemize}
We report results with $k=10$ to facilitate comparison with existing methods.
% By incorporating these metrics, we demonstrate the PCGNet model's capability to capture user personal preferences and provide accurate recommendations, showing a comprehensive evaluation of its effectiveness.

\subsection{Comparison with SOTA Method (RQ.1)}
The overall performance of all compared methods on two datasets
are listed in Table \ref{tab: overall}. We have the following observations:
\begin{enumerate}[leftmargin=*, label={(\arabic*)}]
    \item Our proposed PCGNet establishes new state-of-the-art performance, outperforming other baselines. These gains stem from: a) correlation-aware and context-aware interaction modeling through heterogeneous graph disentanglement, dynamically coordinating visual-textual compatibility with personalized preference signals; b) robust representation learning via share/specific information maximization, suppressing spurious correlations while preserving contextual coherence.
    
    \item Single-modality based methods (MF-BPR, V-BPR, T-BPR) underperform multi-modal approaches across all metrics, with VT-BPR which combines both visual and textual information, improving prediction accuracy over its uni-modal counterparts on both datasets. This validates the necessity of multi-modal fusion for fashion compatibility modeling.

    \item While GP-BPR and PCE-Net jointly model compatibility and personal preference, their decoupled combination strategy shows constrained gains, for example, GP-BPR achieves only 0.8232 AUC while VT-BPR's achieves 0.8105 on Polyvore-U-519, suggesting insufficient capture of overlap interactions between the two objectives.
    
    \item Although CP-TransMatch employs graph-based translation operations (TransE) for triplet user-top-bottom relationships, it achieves suboptimal performance  based on most metrics evaluation, due to oversimplified edge semantics. This highlights the superiority of our dual information maximization strategy in disentangling shared structure information rules and unique compatibility/preference patterns.
\end{enumerate}

\subsection{Ablation experiments (RQ. 2)}
To verify the effectiveness of each module of PCGNet, we designed four variants and compared their recommendation performance against PCGNet.

% \begin{table*}[!t]
% \caption{Performance of PCGNet and and its variants. -w/o- represents removing related components.}
% \begin{center}
% \setlength{\tabcolsep}{3mm}
% \begin{tabular}{l|cccc|cccc}
% \toprule
% \multirow{2}{*}{Method}&\multicolumn{4}{c|}{Polyvore-U-519} & \multicolumn{4}{c}{IQON3000} \\
% \cmidrule(lr){2-5} \cmidrule(lr){6-9}
% & AUC$\uparrow$ & HR@10$\uparrow$ & NDCG@10$\uparrow$ & MRR@10$\uparrow$ & AUC$\uparrow$ & HR@10$\uparrow$ & NDCG@10$\uparrow$ & MRR@10$\uparrow$ \\
% \midrule
% PCGNet-w/o-$\mathcal{L}_\cap$  & 0.9524 & 0.7822 & 0.5910 &0.5305 &0.9580 &0.8010 &0.6275 &0.5725 \\
% PCGNet-w/o-$\mathcal{L}_\varnothing$  & 0.9659 & 0.8359 &0.7084  &0.6599 &0.9670 &0.8265 &0.6712 &0.6225 \\
% % PCGNet-w/o-Aug.  &0.9640  &0.8303  &0.6899  &0.6453 & & & & \\
% PCGNet-w/o-V &0.8894  &0.5915  &0.3490  &0.2736 &0.8810 &0.7125 &0.4992 &0.4323 \\
% PCGNet-w/o-T & 0.9587 &0.8037  &0.6325  &0.5783 &0.9650 &0.8428 &0.7130 &0.6719 \\
% \midrule
% \textbf{PCGNet} & \textbf{0.9667} & \textbf{0.8814} & \textbf{0.7117} & \textbf{0.6704} & \textbf{0.9725} & \textbf{0.9129} & \textbf{0.7386} & \textbf{0.6802}  \\
% \bottomrule
% \end{tabular}
% \end{center}
% \label{tab: ablation}
% \end{table*}

\begin{table*}[!t]
\caption{Performance of PCGNet and and its variants. {\textbf{Bold}} numbers indicate the best results.}
\begin{center}
\setlength{\tabcolsep}{6mm}
\begin{tabular}{ccccc|ccccc}
\toprule
\multicolumn{5}{c|}{Polyvore-U-519} & \multicolumn{5}{c}{IQON3000} \\
% \cmidrule(lr){2-5} \cmidrule(lr){6-9}
\midrule 
V & T & $\mathcal{L}_\cap$ & $\mathcal{L}_\varnothing$ & AUC$\uparrow$ &
V & T & $\mathcal{L}_\cap$ & $\mathcal{L}_\varnothing$ & AUC$\uparrow$ \\
% & AUC$\uparrow$ & HR@10$\uparrow$ & NDCG@10$\uparrow$ & MRR@10$\uparrow$ & AUC$\uparrow$ & HR@10$\uparrow$ & NDCG@10$\uparrow$ & MRR@10$\uparrow$ \\
\midrule
$\times$ & \checkmark & \checkmark & \checkmark & \textit{0.8894}
& $\times$ & \checkmark & \checkmark & \checkmark & \textit{0.8810}\\
 \checkmark & $\times$ & \checkmark & \checkmark & 0.9587 
& \checkmark & $\times$ & \checkmark & \checkmark &  0.9650 \\
 \checkmark & \checkmark & $\times$ & \checkmark  & 0.9524
& \checkmark & \checkmark & $\times$ & \checkmark  & 0.9580 \\
 \checkmark  & \checkmark & \checkmark & $\times$ & 0.9659 
& \checkmark  & \checkmark & \checkmark & $\times$  & 0.9703\\
\midrule
\checkmark  & \checkmark & \checkmark & \checkmark & \textbf{0.9667}
& \checkmark  & \checkmark & \checkmark  & \checkmark & \textbf{0.9725}
\\
\bottomrule
\end{tabular}
\end{center}
\label{tab: ablation}
\end{table*}

\textit{Effectiveness of loss components.} To demonstrate the effectiveness of maximizing view-shared and view-specific product-compatibility$\slash$ personal preference information, we designed two variants (without $\mathcal{L}_\cap$ and without $\mathcal{L}_\varnothing$). Table \ref{tab: ablation} shows the necessity of each component. From the ablation results, we find that it has a greater impact by removing $\mathcal{L}_\cap$ than by removing $\mathcal{L}_\varnothing$.
This can be explained by the fact that optimizing $\mathcal{L}_\cap$ actually maximizes the overall task-relevant information of the shared view, which is more beneficial in our special personalized fashion matching task. However, focusing on view-specific aspects is also crucial to improving the overall recommendation quality.

\textit{Effectiveness of multi-modality.} To assess the impact of different modalities in personalized fashion matching, we further evaluate PCGNet by removing either visual (V) or textual (T) features. As shown in Table \ref{tab: ablation}, the performance declines significantly when visual features are excluded, compared to the removal of textual modality. This is expected, as visual information captures fine-grained details such as color harmony, pattern matching, and overall style coherence, which are difficult to convey solely through textual descriptions. This finding aligns with prior studies in fashion recommendation~\cite{VBPR, GPBPR}, which underscores the dominance of visual signals in compatibility modeling.
On the other hand, the smaller performance drop observed when removing textual features suggests that textual descriptions primarily serve as complementary information. This could be attributed to the fact that textual data often contains noise (e.g., inconsistent product descriptions) and lacks the richness of visual representations. Nevertheless, PCGNet outperforms the other variants, demonstrating that textual features still contribute to advancing model's performance by providing richer contextual information, aiding in disambiguation of personal preferences.

\subsection{Hyperparameter and Other Analyses (RQ. 3)}
\subsubsection{Effect of key hyperparameters }
\begin{figure}[t]
  \centering
  \setlength{\abovecaptionskip}{0.2cm}
  \includegraphics[width=\linewidth]{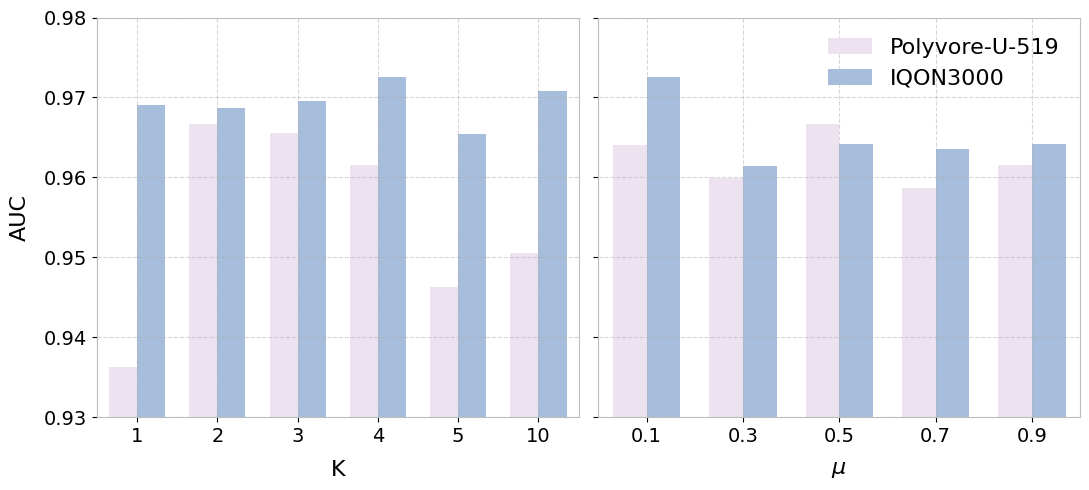}
  \caption{Performance of PCGNet with respect to the different key parameters.}
  % \vspace{-10px}
  \label{fig: hyper}
  % \vspace{-5px}
\end{figure}

Fig. \ref{fig: hyper} illustrates the performance of different hyperparameters (i.e., top $k$ neighbors, $\mu$) in our PCGNet model. For Polyvore-U-519 dataset, the AUC peaks at $k$=2 and gradually decreases as $k$ increases, indicating that a smaller $k$ value is optimal for this dataset. The performance declines at higher $k$ values (e.g., $k$=5 and $k$=10), indicating potential noise from less relevant neighbors. Comparatively, the performance of the model is less sensitive to $k$ in IQON3000 dataset, possibly because of the inherent characteristics of the dataset. Furthermore, we analyzed the variation of the weight parameter $\mu$ with AUC in Eq. (\ref{eq: loss_all}), which serves as a crucial factor in determining the relative importance of adopting maximum share and specific information in our PCGNet. The optimal performance is achieved when $\mu=0.5$ for Polyvore-U-519, and $\mu=0.1$ for IQON3000. This indicates that it is crucial to introduce $\mathcal{L}_\cap$ and $\mathcal{L}_\varnothing$ in our training process.

\subsubsection{Robustness Analysis}
\begin{figure}[t]
  \centering
  % \vspace{-10px}
  \setlength{\abovecaptionskip}{0.2cm}
  \includegraphics[width=\linewidth]{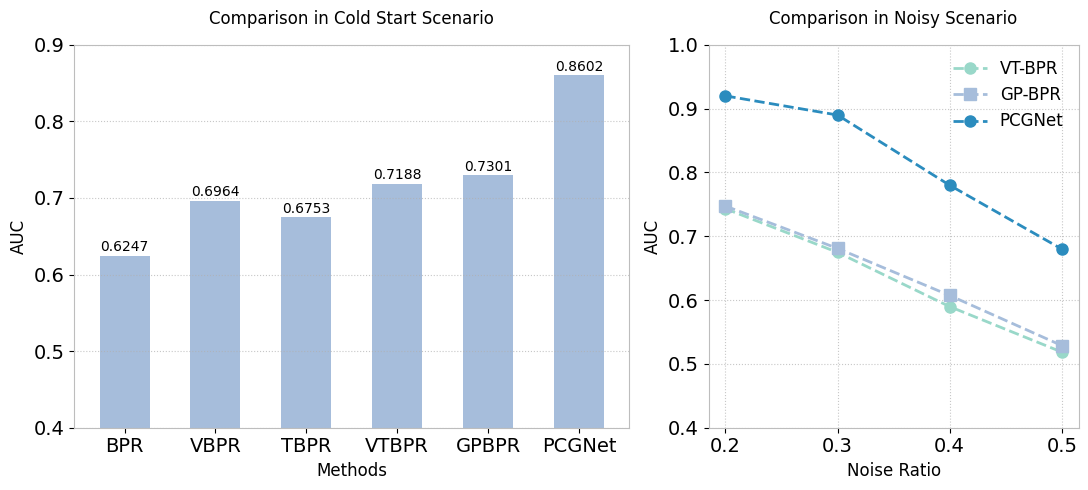}
  % \caption{Performance comparison in cold start environment. }
  \caption{Performance comparison in cold start and noisy environment. }
  \label{fig: cold}
  % \vspace{-10px}
\end{figure}

In a robustness analysis, we specifically focused on evaluating the model's ability to handle cold start and noisy scenarios. On one hand, cold start is known for the inherent challenges due to the lack of historical data for new users or products. We first simulated cold start scenario by isolating a subset of users and products with minimal interaction history. This subset was then used to test the model's ability to make accurate recommendations without relying on extensive past interactions. The results in Fig. \ref{fig: cold} shows that our model significantly outperform other baseline models in cold start scenarios, highlighting the model's enhanced ability to distinguish between relevant and irrelevant products. On the other hand, to evaluate the robustness of PCGNet against structure noise, we perturb the model on the Polyvore-U-519 dataset by randomly adding fake edges. It is evident from Fig. \ref{fig: cold} that with increasing rates of edge perturbing, the performance deteriorates, but still exhibits well robustness compared to other baselines.
% To assess the robustness of PCGNet to structural noise and sparse interaction environments, we perturbed each graph by randomly adding or removing edges. We compared PCGNet to various baselines. As can be seen from the figure, the performance of each method degrades as the edge perturbation rate increases. Notably, our proposed PCGNet outperforms all other methods in both experimental settings, especially in the case of sparse interaction settings.

\subsection{Qualitative Case Study}
\begin{figure}[h]
  \centering
  \includegraphics[width=0.9\linewidth]{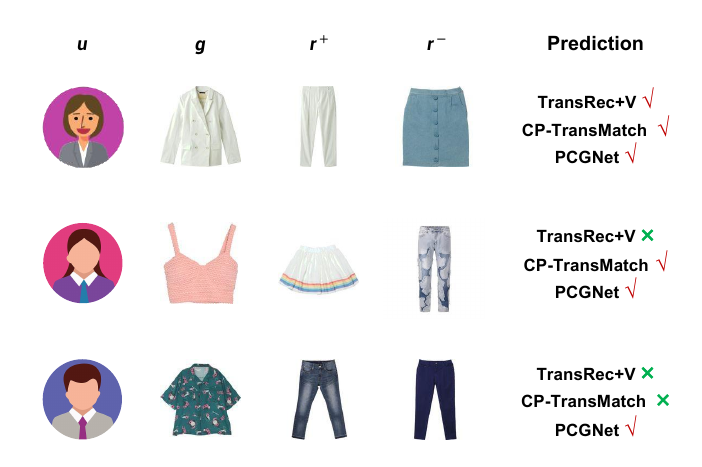}
  \caption{Performance comparison of fashion matching recommendation.}
  \label{fig: case}
  % \vspace{-8px}
\end{figure}
Fig. \ref{fig: case} presents additional comparison case results regarding the precision of fashion matching recommendations. Specifically, we randomly sampled user-top-bottom transaction triplets, denoted as $< u, g, r_+ >$ where a negative product $r_-$ was selected from products that the user had not previously interacted with. We set the user $u$ and the top $t$ as query, and the task needs to predict the positive matching bottom that the user has really chosen. Both top and bottom clothing were represented using visual images and textual descriptions, while user identities were anonymized, retaining only ID information for privacy protection. As shown in Fig. \ref{fig: case}, our PCGNet demonstrates strong performance across all cases. Furthermore, we observed that matching predictions are more straightforward when the product pairs share aesthetic similarities, suggesting that distinct and recognizable features enhance the matching process.

\section{Conclusions and Future Work}
% In this work, we propose a novel framework for personalized fashion matching by leveraging a multiplex graph structure that integrates both product compatibility and user preference views. By unifying the shared and unique information from product compatibility and user personal preference view set across, we effectively capture the latent preference from the triplet relations. The learned representations enable accurate recommendations, demonstrating the potential applications of graph structure learning in fashion recommendation tasks. Experiment results validate the effectiveness of our approach, outperforming traditional methods. 
In this paper, we present PCGNet, a novel framework that enables more precise modeling of customer satisfaction patterns, overcoming the drawbacks of existing fashion recommendation methods.
% that overlook complex interactions between personal preferences and product compatibility. 
By constructing heterogeneous graphs from multiple views and maximizing mutual information between them, PCGNet effectively integrates personalized preferences and compatibility while filtering out task-irrelevant noises. Our approach introduces several key advancements. A multi-objective optimization framework that jointly learns compatibility relationships and personal preferences while preserving their distinct characteristics through view-specific information retention.
% The introduced correlation-aware neighbor sampling and learnable global graph augmentation further refine graph structures so as to capture reliable relationships. 
Algorithmically robust components, including correlation-aware neighbor sampling and learnable graph augmentation, which mitigate data sparsity and noise in real-world fashion datasets. Extensive experiments on benchmark datasets have demonstrated PCGNet's superior recommendation accuracy and robustness compared to state-of-the-art methods, providing a practical solution for enhancing customer satisfaction in fashion e-commerce platforms.

% Despite its strengths, our framework has certain limitations which need further improvement. 
As our future work, certain areas of PCGNet warrant further exploration.
For instance, the mutual information module, functioning as an unsupervised component for multiplex graph learning, primarily provides generic self-supervised signals and my may struggle to capture task-specific supervisory signals (e.g., fine-grained personal preferences). Future research may explore mining latent task-oriented signals, such as integrating hierarchical reconstruction from deeper graph decoder layers, to enhance supervision, or introducing auxiliary supervised tasks to complement the current mutual information objectives.

\begin{acks}
The work described in this paper was supported, in part, by the Innovation and Technology Fund (Project: ITP/004/24TP), The Hong Kong University of Science and Technology (Grant: R9973) and by the Research Institute for Intelligent Wearable Systems (Grant: CD95/ P0049355) of The Hong Kong Polytechnic University. 
\end{acks}

%%
%% The next two lines define the bibliography style to be used, and
%% the bibliography file.
\bibliographystyle{ACM-Reference-Format}
\bibliography{base}

@String{Computing = "Computing" }

@String{Computer = "{IEEE} Computer" }

@String{Springer = "Springer-Verlag" }

@ArtifactSoftware{R,
    title = {R: A Language and Environment for Statistical Computing},
    author = {{R Core Team}},
    organization = {R Foundation for Statistical Computing},
    address = {Vienna, Austria},
    year = {2019},
    url = {https://www.R-project.org/},
}

@inproceedings{VBPR,
  title={VBPR: visual bayesian personalized ranking from implicit feedback},
  author={He, Ruining and McAuley, Julian},
  booktitle={Proceedings of the AAAI conference on artificial intelligence},
  volume={30},
  number={1},
  year={2016}
}

@article{jing2023multimodal,
  title={Multimodal high-order relationship inference network for fashion compatibility modeling in internet of multimedia things},
  author={Jing, Peiguang and Cui, Kai and Zhang, Jing and Li, Yun and Su, Yuting},
  journal={IEEE Internet of Things Journal},
  volume={11},
  number={1},
  pages={353--365},
  year={2023},
  publisher={IEEE}
}

@inproceedings{ECCV2018,
  title={Learning type-aware embeddings for fashion compatibility},
  author={Vasileva, Mariya I and Plummer, Bryan A and Dusad, Krishna and Rajpal, Shreya and Kumar, Ranjitha and Forsyth, David},
  booktitle={Proceedings of the European conference on computer vision (ECCV)},
  pages={390--405},
  year={2018}
}

@inproceedings{NGCF2019,
  title={Neural graph collaborative filtering},
  author={Wang, Xiang and He, Xiangnan and Wang, Meng and Feng, Fuli and Chua, Tat-Seng},
  booktitle={Proceedings of the 42nd international ACM SIGIR conference on Research and development in Information Retrieval},
  pages={165--174},
  year={2019}
}

@article{lu2021outfit,
  title={Outfit compatibility prediction with multi-layered feature fusion network},
  author={Lu, Shufang and Zhu, Xiang and Wu, Yingying and Wan, Xianmei and Gao, Fei},
  journal={Pattern Recognition Letters},
  volume={147},
  pages={150--156},
  year={2021},
  publisher={Elsevier}
}

@article{li2022disentangled,
  title={Disentangled graph neural networks for session-based recommendation},
  author={Li, Ansong and Cheng, Zhiyong and Liu, Fan and Gao, Zan and Guan, Weili and Peng, Yuxin},
  journal={IEEE Transactions on Knowledge and Data Engineering},
  volume={35},
  number={8},
  pages={7870--7882},
  year={2022},
  publisher={IEEE}
}

@article{graphsurvey,
  title={Graph neural networks in recommender systems: a survey},
  author={Wu, Shiwen and Sun, Fei and Zhang, Wentao and Xie, Xu and Cui, Bin},
  journal={ACM Computing Surveys},
  volume={55},
  number={5},
  pages={1--37},
  year={2022},
  publisher={ACM New York, NY}
}

@inproceedings{fan2019graph,
  title={Graph neural networks for social recommendation},
  author={Fan, Wenqi and Ma, Yao and Li, Qing and He, Yuan and Zhao, Eric and Tang, Jiliang and Yin, Dawei},
  booktitle={The world wide web conference},
  pages={417--426},
  year={2019}
}

@inproceedings{jin2020graph,
  title={Graph structure learning for robust graph neural networks},
  author={Jin, Wei and Ma, Yao and Liu, Xiaorui and Tang, Xianfeng and Wang, Suhang and Tang, Jiliang},
  booktitle={Proceedings of the 26th ACM SIGKDD international conference on knowledge discovery \& data mining},
  pages={66--74},
  year={2020}
}

@article{liu2022rgcf,
  title={RGCF: Refined graph convolution collaborative filtering with concise and expressive embedding},
  author={Liu, Kang and Xue, Feng and Hong, Richang},
  journal={Intelligent Data Analysis},
  volume={26},
  number={2},
  pages={427--445},
  year={2022},
  publisher={IOS Press}
}

@article{velivckovic2018deep,
  title={Deep graph infomax},
  author={Veli{\v{c}}kovi{\'c}, Petar and Fedus, William and Hamilton, William L and Li{\`o}, Pietro and Bengio, Yoshua and Hjelm, R Devon},
  journal={arXiv preprint arXiv:1809.10341},
  year={2018}
}

@article{liu2024unifying,
  title={Unifying heterogeneous and homogeneous relations for personalized compatibility modeling},
  author={Liu, Jinhuan and Hou, Lei and Yu, Xu and Song, Xuemeng and Ren, Zhaochun},
  journal={Knowledge-Based Systems},
  volume={290},
  pages={111560},
  year={2024},
  publisher={Elsevier}
}

@article{2024beyond,
  title={Beyond Redundancy: Information-aware Unsupervised Multiplex Graph Structure Learning},
  author={Shen, Zhixiang and Wang, Shuo and Kang, Zhao},
  journal={arXiv preprint arXiv:2409.17386},
  year={2024}
}

@inproceedings{guan2022personalized,
  title={Personalized fashion compatibility modeling via metapath-guided heterogeneous graph learning},
  author={Guan, Weili and Jiao, Fangkai and Song, Xuemeng and Wen, Haokun and Yeh, Chung-Hsing and Chang, Xiaojun},
  booktitle={Proceedings of the 45th international ACM SIGIR conference on research and development in information retrieval},
  pages={482--491},
  year={2022}
}

@inproceedings{transmatch,
  title={Modeling Multi-Relational Connectivity for Personalized Fashion Matching},
  author={Ding, Yujuan and Mok, PY and Bin, Yi and Yang, Xun and Cheng, Zhiyong},
  booktitle={Proceedings of the 31st ACM International Conference on Multimedia},
  pages={7047--7055},
  year={2023}
}

@inproceedings{GPBPR,
  title={GP-BPR: Personalized compatibility modeling for clothing matching},
  author={Song, Xuemeng and Han, Xianjing and Li, Yunkai and Chen, Jingyuan and Xu, Xin-Shun and Nie, Liqiang},
  booktitle={Proceedings of the 27th ACM international conference on multimedia},
  pages={320--328},
  year={2019}
}

@article{li2017mining,
  title={Mining fashion outfit composition using an end-to-end deep learning approach on set data},
  author={Li, Yuncheng and Cao, Liangliang and Zhu, Jiang and Luo, Jiebo},
  journal={IEEE Transactions on Multimedia},
  volume={19},
  number={8},
  pages={1946--1955},
  year={2017},
  publisher={IEEE}
}

@inproceedings{han2017learning,
  title={Learning fashion compatibility with bidirectional lstms},
  author={Han, Xintong and Wu, Zuxuan and Jiang, Yu-Gang and Davis, Larry S},
  booktitle={Proceedings of the 25th ACM international conference on Multimedia},
  pages={1078--1086},
  year={2017}
}

@inproceedings{cui2019dressing,
  title={Dressing as a whole: Outfit compatibility learning based on node-wise graph neural networks},
  author={Cui, Zeyu and Li, Zekun and Wu, Shu and Zhang, Xiao-Yu and Wang, Liang},
  booktitle={The world wide web conference},
  pages={307--317},
  year={2019}
}

@inproceedings{vivek2023personalized,
  title={Personalized Outfit Compatibility Prediction Using Outfit Graph Network},
  author={Vivek, BS and Bhattacharya, Gaurab and Gubbi, Jayavardhana and Pal, Arpan and Balamuralidhar, P and others},
  booktitle={2023 International Joint Conference on Neural Networks (IJCNN)},
  pages={1--8},
  year={2023},
  organization={IEEE}
}

@article{jing2023category,
  title={Category-aware multimodal attention network for fashion compatibility modeling},
  author={Jing, Peiguang and Cui, Kai and Guan, Weili and Nie, Liqiang and Su, Yuting},
  journal={IEEE Transactions on Multimedia},
  volume={25},
  pages={9120--9131},
  year={2023},
  publisher={IEEE}
}

@article{mo2023towards,
  title={Towards private stylists via personalized compatibility learning},
  author={Mo, Dongmei and Zou, Xingxing and Pang, Kaicheng and Wong, Wai Keung},
  journal={Expert Systems with Applications},
  volume={219},
  pages={119632},
  year={2023},
  publisher={Elsevier}
}

@inproceedings{CP,
  title={Modeling Multi-Relational Connectivity for Personalized Fashion Matching},
  author={Ding, Yujuan and Mok, PY and Bin, Yi and Yang, Xun and Cheng, Zhiyong},
  booktitle={Proceedings of the 31st ACM International Conference on Multimedia},
  pages={7047--7055},
  year={2023}
}

@inproceedings{binary,
  title={Learning binary code for personalized fashion recommendation},
  author={Lu, Zhi and Hu, Yang and Jiang, Yunchao and Chen, Yan and Zeng, Bing},
  booktitle={Proceedings of the IEEE/CVF conference on computer vision and pattern recognition},
  pages={10562--10570},
  year={2019}
}

@article{ADAM,
  title={Adam: A method for stochastic optimization},
  author={Kingma, Diederik P},
  journal={arXiv preprint arXiv:1412.6980},
  year={2014}
}

@article{BPR,
  title={BPR: Bayesian personalized ranking from implicit feedback},
  author={Rendle, Steffen and Freudenthaler, Christoph and Gantner, Zeno and Schmidt-Thieme, Lars},
  journal={arXiv preprint arXiv:1205.2618},
  year={2012}
}

@article{PCE,
  title={Attention-Based Personalized Compatibility Learning for Fashion Matching},
  author={Nie, Xiaozhe and Xu, Zhijie and Zhang, Jianqin and Tian, Yu},
  journal={Applied Sciences},
  volume={13},
  number={17},
  pages={9638},
  year={2023},
  publisher={MDPI}
}

@inproceedings{DGSR,
  title={Leveraging two types of global graph for sequential fashion recommendation},
  author={Ding, Yujuan and Ma, Yunshan and Wong, Wai Keung and Chua, Tat-Seng},
  booktitle={Proceedings of the 2021 International Conference on Multimedia Retrieval},
  pages={73--81},
  year={2021}
}

@inproceedings{TransRec,
  title={Translation-based recommendation},
  author={He, Ruining and Kang, Wang-Cheng and McAuley, Julian},
  booktitle={Proceedings of the eleventh ACM conference on recommender systems},
  pages={161--169},
  year={2017}
}

@inproceedings{liao2023,
  title={Recommendation of mix-and-match clothing by modeling indirect personal compatibility},
  author={Liao, Shuiying and Ding, Yujuan and Mok, PY},
  booktitle={Proceedings of the 2023 ACM International Conference on Multimedia Retrieval},
  pages={560--564},
  year={2023}
}

@inproceedings{liao2024reproducibility,
  title={Reproducibility Companion Paper: Recommendation of Mix-and-Match Clothing by Modeling Indirect Personal Compatibility},
  author={Liao, Shuiying and Ding, Yujuan and Mok, PY and Huang, Qiushi and Cao, Jialun},
  booktitle={Proceedings of the 2024 International Conference on Multimedia Retrieval},
  pages={1224--1227},
  year={2024}
}

@article{liao2026consistency,
  title={Consistency regularization for complementary clothing recommendations},
  author={Liao, Shuiying and Mok, PY and Li, Li},
  journal={Applied Soft Computing},
  pages={115069},
  year={2026},
  publisher={Elsevier}
}

@article{liao2026hamiltonian,
  title={Hamiltonian Spectral-Temporal Dissipative Dynamics for Sequential Recommendation},
  author={Liao, Shuiying and Mok, PY},
  journal={arXiv preprint arXiv:2608.25755},
  year={2026}
}

@article{liao2025data,
  title={Data-driven recommendations for fashion- an investigation of personalization with sparse data},
  author={Liao, Shuiying and others},
  year={2025},
  publisher={Hong Kong Polytechnic University}
}

@inproceedings{liao2024hypergraph,
  title={Hypergraph-Enhanced Contrastively Regularized Transformer for Multi-Behavior E-commerce Product Recommendation},
  author={Liao, Shuiying and Mok, PY},
  booktitle={2024 IEEE International Conference on Data Mining (ICDM)},
  pages={767--772},
  year={2024},
  organization={IEEE}
}

@article{yang2023heterogeneous,
  title={A heterogeneous graph neural network model for list recommendation},
  author={Yang, Wenchuan and Li, Jichao and Tan, Suoyi and Tan, Yuejin and Lu, Xin},
  journal={Knowledge-Based Systems},
  volume={277},
  pages={110822},
  year={2023},
  publisher={Elsevier}
}

@article{kipf2016semi,
  title={Semi-supervised classification with graph convolutional networks},
  author={Kipf, Thomas N and Welling, Max},
  journal={arXiv preprint arXiv:1609.02907},
  year={2016}
}

@article{liu2024towards,
  title={Towards self-interpretable graph-level anomaly detection},
  author={Liu, Yixin and Ding, Kaize and Lu, Qinghua and Li, Fuyi and Zhang, Leo Yu and Pan, Shirui},
  journal={Advances in Neural Information Processing Systems},
  volume={36},
  year={2024}
}

@article{maddison2016concrete,
  title={The concrete distribution: A continuous relaxation of discrete random variables},
  author={Maddison, Chris J and Mnih, Andriy and Teh, Yee Whye},
  journal={arXiv preprint arXiv:1611.00712},
  year={2016}
}

@article{jang2016categorical,
  title={Categorical reparameterization with gumbel-softmax},
  author={Jang, Eric and Gu, Shixiang and Poole, Ben},
  journal={arXiv preprint arXiv:1611.01144},
  year={2016}
}

@article{zhang2023constrained,
  title={Constrained Bipartite Graph Learning for Imbalanced Multi-Modal Retrieval},
  author={Zhang, Han and Li, Yiding and Li, Xuelong},
  journal={IEEE Transactions on Multimedia},
  year={2023},
  publisher={IEEE}
}

@article{zhou2022attribute,
  title={Attribute-aware heterogeneous graph network for fashion compatibility prediction},
  author={Zhou, Zhouyi and Su, Zhuo and Wang, Ruomei},
  journal={Neurocomputing},
  volume={495},
  pages={62--74},
  year={2022},
  publisher={Elsevier}
}

@article{guan2022partially,
  title={Partially supervised compatibility modeling},
  author={Guan, Weili and Wen, Haokun and Song, Xuemeng and Wang, Chun and Yeh, Chung-Hsing and Chang, Xiaojun and Nie, Liqiang},
  journal={IEEE Transactions on Image Processing},
  volume={31},
  pages={4733--4745},
  year={2022},
  publisher={IEEE}
}

@incollection{TryonCM2,
  title={Try-On-Enhanced Fashion Compatibility Modeling},
  author={Guan, Weili and Song, Xuemeng and Zhou, Dongliang and Nie, Liqiang},
  booktitle={Advanced Multimodal Compatibility Modeling and Recommendation},
  pages={33--56},
  year={2025},
  publisher={Springer}
}

%%
%% If your work has an appendix, this is the place to put it.
\appendix

\end{document}